\documentstyle[aas2pp4]{article}

\input epsf.sty
\let\simgt=\gtrsim
\let\simlt=\lesssim

\newcommand{\hmpc}{h^{-1}{\rm\,Mpc}}
\newcommand{\impc}{{\rm\,Mpc}^{-1}}
\newcommand{\ihmpc}{h{\rm\,Mpc}^{-1}}

\begin{document}

\title{Can Baryonic Features Produce the Observed 
\boldmath $100h^{-1}$\unboldmath Mpc Clustering?}

\author{Daniel J.\ Eisenstein and Wayne Hu\footnote{Alfred P. Sloan Fellow}}
\affil{Institute for Advanced Study, Princeton, NJ 08540}

\author{Joseph Silk}
\affil{Departments of Astronomy \& Physics, 
UC Berkeley, Berkeley, CA 94720}

\centerline{and}

\author{Alexander S.\ Szalay}
\affil{Department of Physics \& Astronomy, Johns Hopkins University,
Baltimore, MD 21218}


\begin{abstract}
We assess the possibility that baryonic acoustic oscillations in
adiabatic models may explain the observations of excess power in
large-scale structure on $100\hmpc$ scales.  The observed location
restricts models to two extreme areas of parameter space.  In either
case, the baryon fraction must be large ($\Omega_b/\Omega_0 \simgt
0.3$) to yield significant features.  The first region requires
$\Omega_0 \simlt 0.2 h$ to match the location, implying large blue
tilts ($n \simgt 1.4$) 
to satisfy cluster abundance constraints.  The power spectrum also
continues to rise toward larger scales in these models.  The second
region requires $\Omega_0 \approx 1$, implying $\Omega_b$ well out of
the range of big bang nucleosynthesis constraints; moreover, the peak 
is noticeably wider than the observations suggest.
Testable
features of both solutions are that they require moderate reionization
and thereby generate potentially observable $(\sim 1\mu$K) 
large-angle polarization, 
as well as sub-arc-minute temperature fluctuations.  In short, baryonic
features in adiabatic models may explain the observed excess only if
currently favored determinations of cosmological parameters are in
substantial error or if present surveys do not represent a fair sample
of $100\hmpc$ structures.
\end{abstract}

\twocolumn

\section{Introduction}

As the study of large-scale structure has pushed to ever larger scales,
several data samples have suggested the presence of excess power confined 
in a narrow region around the $100\hmpc$ scale.  
The first such claim was the
pencil-beam redshift survey of \cite{Bro90}\ (1990), in which six
concentrations of galaxies separated by a periodic spacing of
$128\hmpc$ were seen.  Later work (e.g.\ \cite{Bah91} 1991; 
\cite{Guz92}\ 1992; \cite{Wil94}\ 1994) has confirmed that these
overdensities are indeed part of extended structures rather than
small-scale anomalies, and new pencil beams show similar behavior
(\cite{Bro95}\ 1995).  More recently, the 2-dimensional power spectrum
of the Las Campanas Redshift Survey (\cite{Lan96}\ 1996, hereafter LCRS)
and the 3-dimensional power spectrum of rich Abell clusters
(\cite{Ein97}\ 1997 and references therein) reveal a narrow peak 
at similar scales, $k\approx0.06\ihmpc$ and $0.052\ihmpc$ respectively.  

Other data sets show anomalies on these scales, although they are
unable to resolve a narrow feature.  Three-dimensional reconstructions
based on the angular correlations in the APM survey (\cite{Gaz97}\ 1997)
suggest a sharp drop in the power spectrum in this region.  Finally,
high-redshift \ion{C}{4} absorption lines in quasar spectra were found
to be correlated on $100\hmpc$ scales (\cite{Qua96}\ 1996); if this is
due to large-scale structure, it indicates greater power than
expected.  Hence, several different lines of observational inquiry
suggest excess power on $100\hmpc$ scales, perhaps in the form of a
narrow peak at wavenumbers $\sim\!0.05h-0.06h \impc$.

Cosmological models based on collisionless dark matter (e.g.\ cold dark
matter), when combined with power-law initial power spectra, produce
smooth power spectra at late times.  Such models therefore cannot match
the feature described above.  However, if baryons are present in
significant quantities, the coupling between them and the cosmic
microwave background photons at redshifts $z\gtrsim1000$ produces 
acoustic oscillations near the $100\hmpc$ scale
(see Eisenstein \& Hu 1997, hereafter EH97, and references therein). 

In this {\it Letter}, we consider whether acoustic features in 
adiabatic models
can explain the narrow peak in the power spectrum at $100\hmpc$ scales.  
This choice
of location for a feature immediately restricts us to two rather extreme
regions of parameter space.  We examine each of these in turn, detailing
their requirements and predictions.

Throughout this paper, $\Omega_0$ is the total density of matter
relative to the critical density; $\Omega_b$ is that of the baryons.
The power-law exponent of the initial power spectrum is denoted $n$
($n=1$ is scale-invariant).  The Hubble constant is written as
$100h{\rm\,km\,s^{-1}\,Mpc^{-1}}$.  We assume a cosmological constant
to make the universe flat; an open universe would have a power spectrum
of identical shape but with a less favorable normalization.

\section{Constraints}

A cosmological model with cold dark matter and baryons exhibits a power
spectrum with a broad global maximum (hereafter, the {\it peak}) at
small wavenumbers $k \simlt 0.05\ihmpc$ and a series of oscillations
(hereafter, the {\it bumps}) at larger wavenumbers $k \simgt
0.05\ihmpc$ (c.f.\ Fig.\ 4).  Therefore, one may either attempt to
associate the peak or the first bump with the observed $100\hmpc$
feature.  This yields two disjoint areas of parameter space which we
display in Figure 1 for two different values of $h$.  We will now
discuss these two regions separately.

\begin{figure}[bt]
\begin{center}
\leavevmode
\epsfxsize=\linewidth \epsfbox{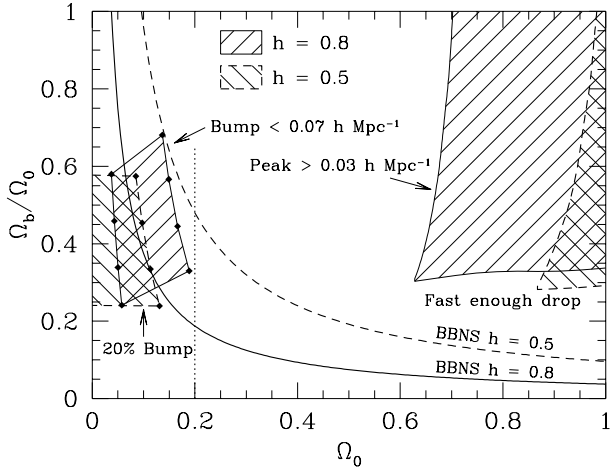}
\end{center}
\caption{Parameter space available for a strong feature at
$k\approx0.05\ihmpc$.  The $\Omega_0$--$\Omega_b/\Omega_0$ plane is
shown for $h=0.5$ (dashed lines) and $h=0.8$ (solid lines).  Left
region: a bump between $0.045h$ and $0.07\ihmpc$ with an amplitude
between 20\% and 160\% as marked by filled dots at 20\%, 40\%, 80\%,
and 160\%.  Right region: the peak at $k \simgt 0.03\ihmpc$ with
an additional requirement on its prominence (positive half-width
half-maximum $\simlt 0.4$ decades [c.f. Fig.~\protect\ref{fig:highomega}]).
Nucleosynthesis constraints of $\Omega_bh^2=0.024$ 
(\protect\cite{Tyt96}\ 1996) are shown (BBNS).}
\label{fig:parameters}
\end{figure}

\subsection{Low-\boldmath$\Omega_0\,$\unboldmath Region}
The region on the left in Figure 1 corresponds to placing the first
bump in the region $0.045\ihmpc<k<0.07\ihmpc$.  The bump shifts to
smaller scales (higher $k$) as $\Omega_0$ increases, as reflected in the
left-right limits.  
We take the bump location to be the position of
corresponding maximum in the oscillatory piece of the transfer function
(\cite{Eis97}, eq.\ 25). 

The lower bound on the baryon fraction
$\Omega_b/\Omega_0$ comes from the requirement that the amplitude of
the bump, as measured using the decomposition of \cite{Eis97}, exceeds
20\% in power.  Smaller oscillations would not explain the
observations.  The upper bound on the baryon fraction comes from
requiring the bump amplitude to be less than a factor of 1.6;
presumably larger oscillations would have caused the second bump at
$k\approx0.12\ihmpc$ to be detected (\cite{Pea94}\ 1994).  Note that
while lowering $h$ from 0.8 to 0.5 causes the allowed region to shift
unfavorably to even lower $\Omega_0$, increasing $h$ to 1.0 only
marginally relaxes the bound on $\Omega_0$.  

\begin{figure}[bt]
\begin{center}
\leavevmode
\epsfxsize=\linewidth 
\epsfbox{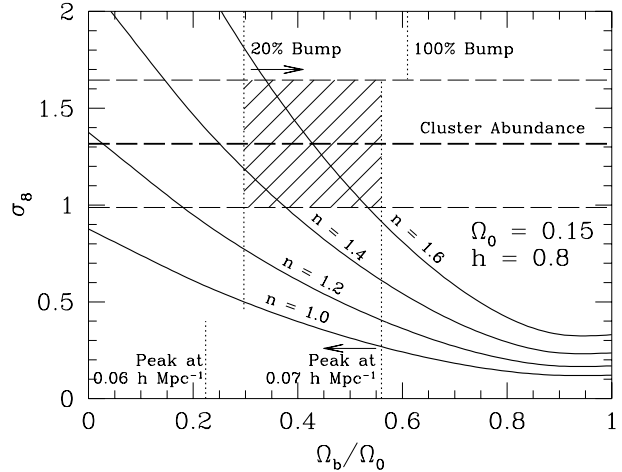}
\end{center}
\caption{The value of $\sigma_8$ is shown as a function of baryon fraction
for a COBE-normalized flat $\Omega_0=0.15$, $h=0.8$ model.  Several 
different values of tilt are shown (solid lines).  The value of $\sigma_8$
required to match the present-day cluster abundance (dashed lines) is taken 
to be $\sigma_8=0.5\Omega_0^{-0.53+0.13\Omega_0}$
(\protect\cite{Eke96}\ 1996) with 25\% variation to reflect errors
here and in the COBE-normalization.  The shaded region represents 
models that satisfy the cluster abundance and bump location while
having $\simgt 20\%$ power enhancement.}
\label{fig:lowomega}
\end{figure}

Hence, one is restricted to a low value of $\Omega_0$, approximately less
than $0.2h$.  For $h\approx0.8$, this does not drastically violate
nucleosynthesis (e.g.\ \cite{Tyt96}\ 1996).  However, the moderate
baryon fraction needed to produce the bumps also causes a significant
suppression of power at $k\gtrsim0.02\ihmpc$.  For a COBE-normalized
(\cite{Bun97}\ 1997) and scale-invariant initial spectrum ($n=1$),
the resulting values of the fluctuations on the cluster scale
$\sigma_8$ are less than 0.5.  This is far smaller than the value
($\gtrsim1.0$ for these $\Omega_0$) needed to reproduce the 
abundance of galaxy clusters. 

Adding a significant blue tilt ($n\gtrsim1.4$) can increase $\sigma_8$
enough to satisfy the cluster abundance constraint.  
We have taken the lowest value in the literature (\cite{Eke96}\ 1996)
to provide conservative lower bounds on $n$.  
We display this situation in Figure 2.
In general, larger amplitude
features must be balanced by larger tilts.  Note that adding a tensor
contribution to COBE or removing the cosmological constant will
decrease the power spectrum normalization and in turn require even
higher tilts (\cite{Whi96}\ 1996).

Tilts of $n\simgt 1.3$ are difficult to realize in inflationary
models (\cite{Bel97}\ 1997).  Empirical constraints 
depend entirely upon the range in
wavenumber used to define the tilt.  The
limited range of scales 
available to {\it COBE} DMR allows only a weak constraint 
($k\approx 10^{-3} \ihmpc$, $n\lesssim1.8$; \cite{Gor96}\ 1996).  Combining
{\it COBE} with degree-scale CMB observations ($k\approx 10^{-2}\ihmpc$) 
limits the tilt more severely.
However both constraints may be
relaxed if the universe were reionized moderately early.
Blue tilts extending to smaller scales are constrained by
arcminute-scale CMB observations ($k \sim 1 \ihmpc$,  $n\lesssim 2$; 
\cite{Vis87} 1987), the absence of spectral distortions from 
dissipation of acoustic waves after thermalization
($k \sim 10^4\impc$, $n\lesssim1.5$; \cite{Hu94}\ 1994), and
limits on primordial black holes ($k \sim 10^{15}\impc$, $n\lesssim 1.3$;
\cite{Gre97} 1997).
In summary, strong blue tilts
that extend from COBE to the smallest observable scales are ruled out,
but between COBE and cluster scales, the situation is less restrictive as
the slope may decrease at smaller scales.

The parameter space remaining to the low-$\Omega_0$ region after the
peak location, peak amplitude, cluster abundance and tilt constraints
are applied is shown as the shaded regions in Figs.~\ref{fig:parameters}
and \ref{fig:lowomega}.

\subsection{High-\boldmath$\Omega_0\,$\unboldmath Region}
The region on the right in Figure 1 corresponds to placing the peak at
$k>0.03\ihmpc$.  Although the lower limit is well below the
observational region $0.05h-0.06\ihmpc$, the peak in these models is
sufficiently broad that the exact maximum need not lie directly on the
preferred scale to yield an enhancement of power.  Figure 1 assumes
$n=1$; adding a blue tilt shifts the peak to higher $k$, increasing the
allowed region.  The region for $n=1.4$ and $h=0.5$ is very similar to
that shown for $h=0.8$.

\begin{figure}[bt]
\begin{center}
\leavevmode
\epsfxsize=\linewidth \epsfbox{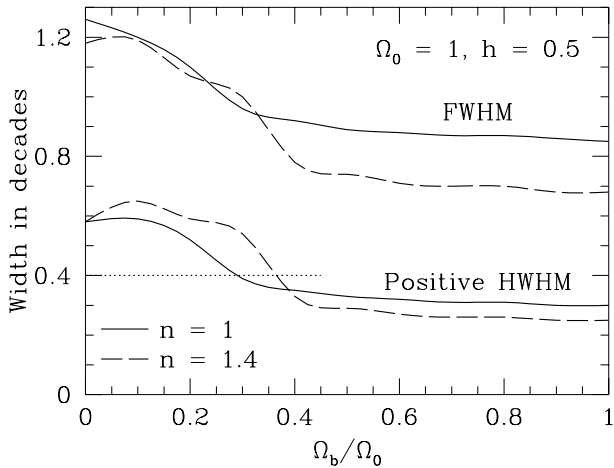}
\end{center}
\caption{The width of the peak of the power spectrum, in decades of
wavenumber, for an $\Omega_0=1$, $h=0.5$ model of varying baryon
fraction.  Displayed are the full-width at half-maximum (FWHM) and the
positive half-width (HWHM), defined as the range in $k$ between the
maximum of the spectrum and its half maximum in the direction of
increasing $k$.  We show $n=1$ models (solid lines) and $n=1.4$ models
(dashed lines).  Increasing $h$ decreases the curves by a small
amount. Requiring a HWHM $\le 0.4$ (dotted lines) eliminates low
baryon models.}
\label{fig:highomega}
\end{figure}

The peak is generically much broader than the bump.  As the
baryon fraction increases, the high-$k$ side of the peak steepens
significantly, giving rise to a prominent and asymmetric feature.  Two
statistics characterizing the width of the peak for an $\Omega_0=1,
h=0.5$ model are shown in Figure 3.
Here one sees that the full-width at half-maximum (FWHM) always exceeds
0.85 decades in $k$ for $n=1$ and 0.65 decades for $n=1.4$.  Similarly
the range in $k$ over which the power spectrum drops from its peak to
its half-maximum in the high-$k$ direction (positive HWHM)
always exceeds 0.3 decades
for $n=1$ and 0.25 decades for $n=1.4$.  Adding a blue tilt steepens
the low-$k$ side of the peak, thereby decreasing the width.

Hence, even for large baryon fractions, the peak may be too broad when
compared with the narrow feature observed in the cluster power
spectrum, LCRS, or pencil-beams.  However, the sharp break on the
small-scale side of the maximum may be a sufficient departure from the
usual low-baryon spectral shape as to allow these models to be
statistically consistent with these observations.  To reflect this
situation, we place a lower limit on the baryon fraction in Figure
\ref{fig:parameters} by requiring that the positive HWHM be less than
0.4 decades in $k$ (c.f.~Fig.~\ref{fig:highomega}).

\section{Discussion}

\begin{figure}[bt]
\begin{center}
\leavevmode
\epsfxsize=\linewidth \epsfbox{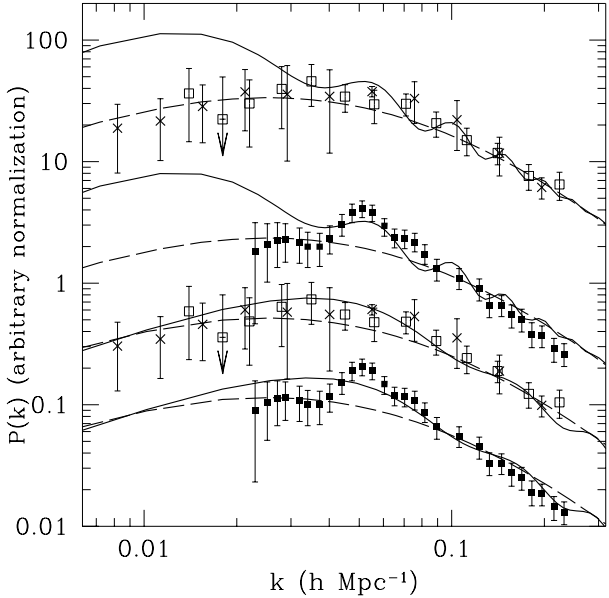}
\end{center}
\caption{Two representative models from the studied regions compared
with large-scale structure data.  The top two solid curves show an
$\Omega_0=0.12$, $\Omega_b=0.04$, $h=0.8$, $n=1.6$ model; the bottom
two solid curves show an $\Omega_0=1$, $\Omega_b=0.4$, $h=0.6$, $n=0.95$
model.  Dashed curves are a $\Gamma\equiv\Omega_0h=0.25$ zero-baryon
model for comparison.  Data sets are APM (\protect\cite{Gaz97}\ 1997)
({\it crosses}), the compilation of \protect\cite{Pea94}\ (1994)
({\it open boxes}), and the cluster power spectrum of
\protect\cite{Ein97}\ (1997) ({\it solid boxes}).  All error bars are
2-$\sigma$, and only data at $k<0.25\ihmpc$ have been plotted.
All normalizations are arbitrary.  The models, if
COBE-normalized, partially reionized, and assumed flat, have $\sigma_8$
of 1.37 (no tensors) and 0.63 (with tensors), respectively.}
\label{fig:poster}
\end{figure}

In Figure 4, we show a representative example from each of the allowed
regions and overlay them with observational data sets.  
The top two curves show an $\Omega_0=0.12$,
$\Omega_b=0.04$, $h=0.8$, $n=1.6$ model.  The first bump is located
near $k=0.06\ihmpc$ and contains significantly more power than a
zero-baryon, $\Gamma\equiv \Omega_0h=0.25$ model (dashed line).  
The first bump is prominent and well-matched to the \cite{Ein97}\ 
(1997) power spectrum; a similar model would fit the LCRS data.
However, the peak at larger scales is yet higher, implying that power should
continue to rise as we look toward larger scales.  This is a generic
feature of this region of parameter space---avoided only by enormous
blue tilts ($n\gtrsim2.3$)---and may well be
incompatible with the turnover in the power spectrum suggested by the
APM survey (\cite{Bau93} 1993; \cite{Gaz97}\ 1997).  Non-linearities
would likely help to wash out the series of bumps at smaller scales
($k\gtrsim0.1\ihmpc$).

The bottom two curves show an $\Omega_0=1$, $\Omega_b=0.4$,
$h=0.6$, $n=0.95$ model.  Again, the model has excess power on $100\hmpc$
scales relative to a $\Gamma=0.25$ model.  
Because of the high baryon fraction, this model
does in fact produce the $\sigma_8$ needed to match the $\Omega_0=1$
cluster abundance.  Of course, the baryon density is in complete
violation of bounds from nucleosynthesis (\cite{Tyt96}\ 1996).
Due to its large width, the peak feature provides only a marginal,
but perhaps adequate, fit to the \cite{Ein97}\ (1997) data.

Although unusual, these models need not be at odds with current CMB
observations.  High baryon fractions tend to substantially enhance the
first acoustic peak, and of course blue tilts enhance all power at
smaller angular scales.  If reionization were not invoked, the models
would overproduce degree-scale anisotropies.  However, with
reionization corresponding to an optical depth of $\tau=0.75$ for the
$\Omega_0=0.12$ model and $\tau=0.5$ for the $\Omega_0=1$ model, the
degree-scale predictions are suppressed to match current observations
(e.g.\ \cite{Net97}\ 1997).  Because of the high baryon content, these
values of $\tau$ correspond to rather low epochs of reionization,
$z_{\rm ri}=33$ and $13$ respectively.  Nor does the reionization
overproduce secondary anisotropies; we find a Vishniac
contribution (\cite{Hu96}\ 1996) across the ATCA band
($\ell\approx4500$) of $\Delta T/T = 2.7 \times 10^{-6}$ and $1.9 \times
10^{-6}$ respectively, well below the current limit of
$1.6 \times 10^{-5}$ (\cite{Sub93}\ 1993).

\begin{figure}[t]
\begin{center}
\leavevmode
\epsfxsize=\linewidth \epsfbox{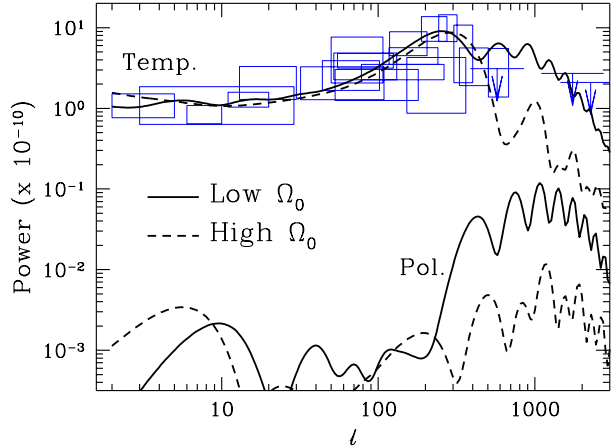}
\end{center}
\caption{CMB temperature anisotropy (including Vishniac effect) 
and polarization
predictions for the two models described in Figure 4.  The models
are consistent with the current observational limits 
(1$\sigma$ error boxes on detections and 2$\sigma$ upper limits,
see e.g. \protect\cite{Smo97}\ 1997 and references therein.)}
\label{fig:cmb}
\end{figure}

Two predictions of these models for the CMB (Fig.\ \ref{fig:cmb}) are
1) that the second acoustic peak will be quite suppressed compared to
the first and third for the high $\Omega_0$ model, due to the high
baryon content, and 2) that the high optical depth will produce
substantial CMB polarization levels, approaching bandpowers of $5
\times 10^{-7}$ at {\it COBE} scales,
as well as substantial sub-arc-minute
temperature fluctuations.  These are within reach of the
current generation of CMB polarization 
experiments (\cite{Kea97}\ 1997) and interferometer experiments,
respectively.

In summary, adiabatic CDM+baryon universes with power-law initial power
spectra produce the peak found
at $k\approx0.05-0.06\ihmpc$ only in extreme regions of
cosmological parameter space. 
Placing the first baryonic bump at these wavenumbers requires
values of $\Omega_0$ lower than those implied by dynamical
mass measurements.  This in turn requires 
extremely large blue tilts and moderate reionization. 
Avoiding tilts above $n\approx 1.7$ necessitates a cosmological constant
that exceeds limits from gravitational lens surveys (\cite{Koc96}\ 1996).
These models may also be in conflict with power spectrum observations at
even larger scales ($k\sim0.02\ihmpc$).  

On the other hand, placing the peak of the power spectrum at the
observed scale requires high values of $\Omega_0$.  Such models need
$\Omega_b \simgt 0.3$ and even so provide a feature that is broader
than the observations suggest.  
Dynamically favored values of $\Omega_0$,
say $\sim\!0.3$, place the first {\it valley} of the power spectrum at
the desired place!

The question remains as to whether the observations fairly sample the
true power spectrum. 
The narrow width of the observed features may merely indicate that a
small number of $k$-modes are dominating the sample.  This is more
likely if the distribution of amplitudes is non-Gaussian; for example,
small nonlinearities in the density field increase the frequency of hot
spots in realizations of the power spectrum (\cite{Ame94} 1994).  If
the underlying theory has a broad peak around $100 \hmpc$, different
volumes may by chance produce spikes at slightly different locations,
with the true width only being recovered in a larger survey.  However,
some underlying feature will still be required, as shown by the failure
of simulations of trace-baryon models to reproduce the observations (LCRS).

Can mildly non-linear evolution shift the location of the peak in the
linear power spectrum?  Second-order corrections to the real-space
power spectrum (e.g. \cite{Jai94} 1994) act only to reduce the
amplitude of features, although the effects are quite small on the
scales in question.  One possible loophole is coherent effects in
redshift space, which we plan to investigate using the Zel'dovich
approximation (\cite{Sza97} 1997).  A second possibility is
scale-dependent bias, for example if objects tend to trace the scale at
which the power spectrum is steepest, rather than where it has its
maximum.

Finally, one may consider models beyond those treated here.
Isocurvature models (e.g.\ \cite{Pee87} 1987) produce a sequence
of oscillations that are $90^\circ$ out of phase with those of
adiabatic models (\cite{Hu96s} 1996; \cite{Sug97} 1997).  
For $\Omega_0\sim 0.3$, this places the peak of the
power spectrum at the intended scale; the first bump is never
relevant.  Alternatively, one can place a feature directly in the
initial power spectrum (\cite{Atr97}\ 1997).  Ongoing redshift surveys
should measure the power spectrum to sufficient precision to
distinguish between these various explanations of the $100\hmpc$
excess.

\noindent{\it Acknowledgments:}
We thank J.R.\ Bond and Uros Seljak for useful discussions,
C.\ Baugh for supplying his results in electronic form, and
the hospitality of the Aspen Center for Physics.
The CMBfast package (\cite{Sel96}\ 1996)
was used to generate numerical transfer functions.
W.H. acknowledges support from the W.M. Keck foundation,
D.J.E.\ and W.H.\ from NSF PHY-9513835,
J.S.\ from NASA and NSF grants, and A.S.Z.\ from a NASA LTSA.

\end{document}